\documentclass[twocolumn,times]{aastex631}

\usepackage{newtxtext,newtxmath}
\usepackage[T1]{fontenc}

\usepackage{nicefrac}
\usepackage{latexsym}
\usepackage{amsmath, mathrsfs, bm}
\usepackage{eqnarray}
\usepackage{url}
\usepackage{cases}
\usepackage{epsfig}
\usepackage{color}
\usepackage{graphicx}
\usepackage{grffile}
\usepackage{float}
\usepackage{soul}
\usepackage{enumerate}
\usepackage{multirow}
\usepackage{threeparttable}
\usepackage{xspace}

\usepackage{bm}
\usepackage{times}

\usepackage{etoolbox}
\preto\align{\par\nobreak\noindent}
\expandafter\preto\csname align*\endcsname{\par\nobreak\noindent}
\preto\multline{\par\nobreak\noindent}
\expandafter\preto\csname align*\endcsname{\par\nobreak\noindent}
\preto\flalign{\par\nobreak\noindent}
\expandafter\preto\csname align*\endcsname{\par\nobreak\noindent}
\preto\eqnarray{\par\nobreak\noindent}
\expandafter\preto\csname align*\endcsname{\par\nobreak\noindent}

\newcommand{\reffg}[1]{Figure~\ref{#1}}
\newcommand{\reftb}[1]{Table~\ref{#1}}
\newcommand{\refeq}[1]{Equation~(\ref{#1})}
\newcommand{\refsc}[1]{Section~\ref{#1}}
\definecolor{indiagreen}{rgb}{0.07, 0.53, 0.03}

\def\hi{\textsc{Hi}\xspace}
\def\hiim{\textsc{Hi\xspace IM}\xspace}

\def\fathomer{FATHOMER\xspace}
\def\crafts{CRAFTS\xspace}

\begin{document}

\title{FAST drift scan survey for \hi intensity mapping: II. stacking-based beam construction of the 19-feed array at $1.4$ GHz}

\author[0009-0008-2564-9398 ]{Xinyang Zhao}
\affiliation{Liaoning Key Laboratory of Cosmology and Astrophysics,  College of Sciences, Northeastern University, Shenyang 110819, China}

\correspondingauthor{Yichao Li}
\author[0000-0003-1962-2013]{Yichao Li}
\email{liyichao@mail.neu.edu.cn}
\affiliation{Liaoning Key Laboratory of Cosmology and Astrophysics,  College of Sciences, Northeastern University, Shenyang 110819, China}

\author[0009-0006-2521-025X]{Wenxiu Yang}
\affiliation{National Astronomical Observatories, Chinese Academy of Sciences, Beijing 100101, China}

\affiliation{School of Astronomy and Space Science, University of Chinese Academy of Sciences, Beijing 100049, China}
\affiliation{Key Laboratory of Radio Astronomy and Technology, Chinese Academy of Sciences, A20 Datun Road, Chaoyang District, Beijing 100101, China}

\author[0000-0001-8075-0909]{Furen Deng}
\affiliation{National Astronomical Observatories, Chinese Academy of Sciences, Beijing 100101, China}
\affiliation{School of Astronomy and Space Science, University of Chinese Academy of Sciences, Beijing 100049, China}
\affiliation{Key Laboratory of Radio Astronomy and Technology, Chinese Academy of Sciences, A20 Datun Road, Chaoyang District, Beijing 100101, China}
\affiliation{Institute of Astronomy, University of Cambridge, Madingley Road, Cambridge, CB3 0HA, UK\\}

\author[0000-0003-0631-568X]{Yougang Wang}
\affiliation{National Astronomical Observatories, Chinese Academy of Sciences, Beijing 100101, China}
\affiliation{Key Laboratory of Radio Astronomy and Technology, Chinese Academy of Sciences, A20 Datun Road, Chaoyang District, Beijing 100101, China}
\affiliation{School of Astronomy and Space Science, University of Chinese Academy of Sciences, Beijing 100049, China}
\affiliation{Liaoning Key Laboratory of Cosmology and Astrophysics,  College of Sciences, Northeastern University, Shenyang 110819, China}

\author[0000-0002-6174-8640]{Fengquan Wu}
\affiliation{National Astronomical Observatories, Chinese Academy of Sciences, Beijing 100101, China}
\affiliation{School of Astronomy and Space Science, University of Chinese Academy of Sciences, Beijing 100049, China}
\affiliation{Key Laboratory of Radio Astronomy and Technology, Chinese Academy of Sciences, A20 Datun Road, Chaoyang District, Beijing 100101, China}

\author[0000-0002-2472-6485]{Xin Wang}
\affiliation{School of Physics and Astronomy, Sun Yat-Sen University, No. 2 Daxue Rd., Zhuhai 519082, People's Republic of China}
\affiliation{CSST Science Center for the Guangdong–Hong Kong–Macau Greater Bay Area, SYSU, People's Republic of China}

\author[0000-0002-3464-5128]{Xiaohui Sun}
\affiliation{School of Physics and Astronomy, Yunnan University, Kunming, 650091, P. R. China}

\author[0000-0002-6029-1933]{Xin Zhang}
\affiliation{Liaoning Key Laboratory of Cosmology and Astrophysics,  College of Sciences, Northeastern University, Shenyang 110819, China}
\affiliation{National Frontiers Science Center for Industrial Intelligence and Systems Optimization, Northeastern University, Shenyang 110819, China}
\affiliation{Key Laboratory of Data Analytics and Optimization for Smart Industry (Ministry of Education), Northeastern University, Shenyang 110819, China}

\author[0000-0001-6475-8863]{Xuelei Chen}
\affiliation{National Astronomical Observatories, Chinese Academy of Sciences, Beijing 100101, China}
\affiliation{Liaoning Key Laboratory of Cosmology and Astrophysics,  College of Sciences, Northeastern University, Shenyang 110819, China}
\affiliation{Key Laboratory of Radio Astronomy and Technology, Chinese Academy of Sciences, A20 Datun Road, Chaoyang District, Beijing 100101, China}
\affiliation{School of Astronomy and Space Science, University of Chinese Academy of Sciences, Beijing 100049, China}

\begin{abstract}

Neutral hydrogen (\hi) intensity mapping (IM) presents great promise 
for future cosmological large-scale structure surveys.
However, a major challenge for \hiim cosmological studies is to 
accurately subtract the foreground contamination. 
An accurate beam model is crucial for improving the quality of foreground subtraction.
In this work, we develop a stacking-based beam reconstruction method
utilizing the radio continuum point sources within the drift-scan field.
Based on the Five-hundred-meter Aperture Spherical radio Telescope (FAST), 
we employ two sets of drift-scan survey data and merge the measurements to
construct the beam patterns of the 19 FAST L-band feeds. 
To model the beams, we utilize the Zernike polynomial (ZP), which effectively captures 
asymmetric features of the main beam and the different side lobes.
Due to the symmetric location of the beams, the main features of the beams 
are closely related to the distance from the center of the feed array, e.g.,
as the distance increases, side lobes become more pronounced.
This modeling pipeline leverages the stable drift-scan data to extract 
beam patterns while accounting for and excluding the reflector's changing effects. 
It provides a more accurate measurement beam and a more precise model beam
for FAST \hiim cosmology surveys.

\end{abstract}

\keywords{ Large-scale structure of the universe (902) --- Astronomical methods (1043) --- Radio astronomy (1338) --- Surveys (1671) -- Drift scan imaging (410)}

\section{Introduction}

Neutral hydrogen intensity mapping (\hiim), which is to measure the total \hi intensity of many galaxies 
within large voxels \citep[e.g.][]{2004MNRAS.355.1339B,2006ApJ...653..815M},
has been proposed as a novel method for cosmic large-scale structure surveys in the epoch of 
post-reionization
\citep{2008PhRvL.100i1303C,2008PhRvL.100p1301L,
2008PhRvD..78b3529M,2008PhRvD..78j3511P,2008MNRAS.383..606W,2008MNRAS.383.1195W,
2009astro2010S.234P,2010MNRAS.407..567B,2010ApJ...721..164S,2011ApJ...741...70L,
2012A&A...540A.129A,2012RPPh...75h6901P,2013MNRAS.434.1239B}.
It can be rapidly conducted and expanded to cover substantial survey areas,
making it well-suited for cosmological surveys
\citep{2015ApJ...798...40X,2020JCAP...03..051J,Zhang:2021yof,Jin:2021pcv,Wu:2021vfz,2023JCAP...06..052W,2023SCPMA..6670413W,2023MNRAS.524.2420Z,2025JCAP...01..080P}.

The \hiim technique was investigated by measuring the cross-correlation function 
between an \hiim survey and an optical galaxy survey. 
\citet{2010Natur.466..463C} first reported a cross-correlation detection using 
the \hiim obtained with the Green Bank Telescope (GBT). 
Subsequent studies have also observed cross-correlation power spectra using various telescopes
\citep{2013ApJ...763L..20M,2018MNRAS.476.3382A,2017MNRAS.464.4938W,2022MNRAS.510.3495W,
2023ApJ...947...16A}.
Recently, the MeerKAT \hiim survey
\citep{2015ApJ...803...21B,2021MNRAS.504..208C,
2021MNRAS.501.4344L,2021MNRAS.505.3698W, 2021MNRAS.505.2039P,2023MNRAS.524.3724C}
reported the cross-correlation power spectrum detection with the
optical galaxy survey \citep{2023MNRAS.518.6262C,2024arXiv240721626M,2024arXiv241206750C}.
Meanwhile, the \hiim auto power spectrum on Mpc scales is detected 
using the MeerKAT interferometric observations \citep{2023arXiv230111943P}.
However, the \hiim auto power spectrum on large scales 
remains undetected \citep{2013MNRAS.434L..46S}.
Besides, several \hiim experiments are targeting the post-reionization epoch, 
with some currently collecting data, including projects like the Tianlai project 
\footnote{\url{https://tianlai.bao.ac.cn}}
\citep{2012IJMPS..12..256C,2020SCPMA..6329862L,2021MNRAS.506.3455W,2022MNRAS.517.4637P,2022RAA....22f5020S}
and the Canadian Hydrogen Intensity Mapping Experiment 
~\citep[CHIME \footnote{\url{https://chime-experiment.ca/en}};][]{2014SPIE.9145E..22B},
while others are in the construction phase, such as 
the Baryonic Acoustic Oscillations from Integrated Neutral Gas Observations
~\citep[BINGO \footnote{\url{https://bingotelescope.org}};][]{2013MNRAS.434.1239B} and the
Hydrogen Intensity and Real-time Analysis eXperiment 
~\citep[HIRAX \footnote{\url{https://hirax.ukzn.ac.za}};][]{2016SPIE.9906E..5XN}.
The \hiim technique is also proposed as the key cosmology project with 
Square Kilometre Array 
~\citep[SKA \footnote{\url{https://www.skao.int}};][]{2015aska.confE..19S,2020PASA...37....7S}.
With the Five-hundred-meter Aperture Spherical radio Telescope,
\citep[FAST;][]{2011IJMPD..20..989N,2016RaSc...51.1060L}
\hiim survey shows considerable potential for cosmological studies
\citep{2017PhRvD..96f3525L,2020MNRAS.493.5854H}.

The major challenge for \hiim cosmology studies is to separate the 
\hi brightness fluctuation from the foreground contamination, 
i.e. the synchrotron or free-free emission of the Milky Way and radio
galaxies, which are typically several orders of magnitude brighter than the
cosmological signal \citep{2015aska.confE..35W,2022MNRAS.509.2048S}.
Exploiting the advantageous characteristic of foreground contamination exhibiting
smoothness across the frequency spectrum, it becomes feasible to be 
separated from the fluctuating HI signal 
\citep{2009ApJ...695..183B,2012A&A...540A.129A}.
However, systematic effects exist during the observation 
significantly increase the degree of freedom of the foreground 
frequency spectrum and challenges the foreground separation technique
\citep{2021MNRAS.506.5075M,2022MNRAS.509.2048S,2024MNRAS.527.4717I}.
The primary beam effect is one of the significant systematic effects that 
must be taken into account while analyzing the \hiim survey data.

The primary beam effect refers to two important aspects of the antenna beam pattern. 
The first is the main lobe shape and its frequency dependence; the second is the side-lobe beam pattern level and its frequency-dependent variation.
Both can weaken the smooth foreground spectrum 
assumption, resulting in a failed foreground subtraction \citep{2021MNRAS.506.5075M,2022MNRAS.509.2048S}.
To address such primary beam effects, the observed \hi maps are usually degraded to a common beam size,
i.e. the largest beam size within the observation bandwidth, with an assumption of simple 
Gaussian primary beam shape \citep{2013ApJ...763L..20M,2018MNRAS.476.3382A,2023MNRAS.518.6262C}.
This approach can reduce the primary beam effect to a subdominant level while also 
significantly increasing the signal-to-noise ratio. 
Recently, a deep-learning-based strategy has been developed with simulated data for reducing the 
\hiim survey systematic effect, which includes both the primary beam effect \citep{2022ApJ...934...83N}
and the polarization leakage effect \citep{2023MNRAS.525.5278G}.
To apply such a strategy to the realistic observation data, an accurate beam model is required for training the deep learning network.

The beam pattern can be properly measured via radio holographic measurements with an 
interferometer \citep{2019MNRAS.485.4107I,2021MNRAS.502.2970A,2024arXiv240800172A}, 
or by scanning a bright celestial calibrator for a single dish telescope \citep[e.g.][]{2020RAA....20...64J}.
Besides, the Tianlai experiment has measured the primary beam via unmanned aerial vehicle \citep{2021IAPM...63f..98Z}
and CHIME has measured the primary beam via the Sun transition \citep{2022ApJ...932..100A}.

In this work, we estimate the Stokes I primary beam patterns of the FAST telescope 
that combine the effects of its reflector and L-band feed array.
The FAST L-band observation uses a 19-feed array mounted
in the focus cabin suspended on the cables about the reflector. The reflector consists
of about $4\,400$ active reflector panels, which form a $300$ m paraboloid at each pointing
direction. Such a complicated reflector and feed arrangement heavily increases beam pattern 
instabilities, resulting in a direction-dependent beam pattern \citep{2013PASA...30...41D}.
In this work, we developed a stacking-based beam pattern construction method employing 
\hiim drift scan observation data itself. Because the beam shape is measured in the same
direction as the \hiim survey observation, we may minimize the direction dependency impact
and achieve the beam pattern that is close to the actual case during the observation.

The rest of the paper is organized as follows:
\refsc{sec:data} introduces the datasets used in this work. 
In \refsc{sec:method}, we describe the detailed stacking-based beam pattern reconstruction method. 
The results and discussion are presented in \refsc{sec:results}, 
followed by our conclusions in \refsc{sec:conclusion}.

\section{Data and preprocessing} \label{sec:data}


\begin{table}
\begin{center}
\caption{The observational data used in this work.
In these observations, the low-level noise diode
is adopted and the feed array is rotated to a fixed angle of $23.4^\circ$ 
during each observation to have an optimized sky coverage \citep{2018IMMag..19..112L}.
}\label{tb:data}
\begin{threeparttable}
 \begin{tabular}{ccccc} 
 \hline \hline 
 \multicolumn{2}{c}{\fathomer\tnote{$\dagger$}} &  & \multicolumn{2}{c}{\crafts\tnote{$\ddagger$}} \\ 
   Field center\tnote{$\star$}&Date &       &Field center & Date \\  
   \cline{1-2} \cline{4-5}
     1100+2600&20210302&   &0330+2715&20220418\\
     1100+2600&20220210&   &0330+2737&20201029\\
     1100+2610&20210309&   &0330+2759&20201118\\
     1100+2610&20210314&   &0330+2820&20201027\\
     1100+2610&20220211&   &0330+2904&20200805\\
     1100+2621&20210313&   &0330+2925&20200806\\
     1100+2621&20220212&   &0330+2947&20200821\\
     1100+2632&20210305&   &0330+3009&20200822\\
     1100+2632&20220213&   &0330+3030&20200810\\
     1100+2643&20210306&   &0330+3052&20200813\\
     1100+2643&20220214&   &0330+3114&20200818\\
     1100+2654&20210307&   &0330+3135&20220330\\
     1100+2654&20220215&   &0330+3157&20220324\\
     1100+2705&20220216&   &0330+3219&20220328\\
     1100+2715&20220222&   &         &        \\
     1100+2726&20220217&   &         &        \\
     1100+2737&20220218&   &         &        \\
     1100+2748&20220219&   &         &        \\ \hline
 \end{tabular}
 \begin{tablenotes}
 \scriptsize
 \item[$\dagger$] The \fathomer observations are carried out for $4$ hours drift-scan every night with $1$ second integration time. 
 A low-level noise diode signal is injected for $1\rm s$ in every $8\,{\rm s}$.
 \item[$\ddagger$] The \crafts observations are carried out for $5$ hours drift-scan every nights with $0.2$ second integration time. 
 A low-level noise diode signal is injected with high-cadence,
 injected for $81.92\mu {\rm s}$ in every $198.608\,{\rm \mu s}$..
 \item[$\star$] The field center, encoded with `HHMM+ddmm', indicates the center of each drift scan survey field, where `HHMM' represents the R.A. and `+ddmm' represents the declination, respectively.
 \end{tablenotes}
\end{threeparttable}
\end{center}
\end{table}

\subsection{\fathomer}

We employ the drift scan observation data from the
FAst neuTral HydrOgen intensity Mapping ExpeRiment \citep[\fathomer;][]{2023ApJ...954..139L}.
In this analysis, $18 \times 4$ hr drift scan observation data 
with integration time of $1$ s per timestamp are used, which targets the area 
that covers the R.A. range from 9 to 13 hr, overlapping with the Northern Galactic Cap (NCP) area of the 
Sloan Digital Sky Survey \citep[SDSS;][]{2016MNRAS.455.1553R}.
To achieve in-time calibration, a noise diode signal with known power is injected for $1\,{\rm s}$ in every $8\,{\rm s}$.
We adopt a low-level noise diode
\footnote{The FAST flux calibration uses a noise diode with two power levels: 
$\sim 1\,{\rm K}$ (low-level) and $\sim 10\,{\rm K}$ (high-level). 
For both FATHOMER and CRAFTS drift scan observations, the low-level noise diode injection is used.} 
during the observations.
The drift scan time-ordered data are preprocessed via the analysis pipeline developed in 
\cite{2023ApJ...954..139L}. Following the pipeline, the full frequency band is first split into 
three sub-bands, i.e. the low-frequency band $1050$--$1150$ MHz, the mid-frequency band $1150$--$1250$ MHz, 
and the high-frequency band $1250$--$1450$ MHz. 
In this work, we use the high-frequency band data as it is less contaminated by RFI.
The high-frequency band also covers the continuum spectrum range of the celestial point source catalog,
which is used as a sky model for beam pattern construction.
Extending the lower frequency band requires a better knowledge of the source spectrum model.
The selected data are then RFI flagged and flux calibrated before proceeding to the analysis of this work.

\subsection{CRAFTS}

The Commensal Radio Astronomy FasT Survey \citep[CRAFTS;][]{2018IMMag..19..112L}
is designed to cover the FAST sky between declination of $-14^\circ$ and $+66^\circ$
and proposed to carry out multiple surveys simultaneously in a single drift scan observation,
including Galactic \hi, extra-galactic \hi, pulsar search, and transient surveys.
This analysis uses $14 \times 5$ hr CRAFTS drift scan observation data, 
which targets the area 
that covers the R.A. range from $1$ to $6$ hr and the declination range from  $27^\circ15'$ to $32^\circ19'$.  
CRAFTS drift scan observation data have the high-cadence noise diode 
signal injection\footnote{
To avoid data contamination in Fourier space for pulsar search, the noise injection period for CRAFTS
is set to $198.608\mu {\rm s}$ with $41\%$ duty cycle, i.e. with $81.92\mu {\rm s}$ noise diode on
in every injection period \citep{2024arXiv241208173Y}.}
and the noise diode temperature was pre-calibrated \citep{2020RAA....20...64J}.
The pre-pipeline of CRAFTS data analysis can be found in \citet{2024arXiv241208173Y}.
The integration time for \crafts is $0.2$ s per timestamp. We reduce the integration 
time to $1$ s after the calibration processing. 
We adopt the same frequency partition strategy and time-ordered data 
analysis pipeline as the FATHOMS analysis.
All the FAST drift scan data used in this work are summarized in \reftb{tb:data}.

\subsection{NVSS}

Our stacking-based beam construction requires a sky model, which is constructed using the 
continuum flux density measurements from the NRAO-VLA Sky Survey (NVSS) catalog \citep{2008AJ....136..684K}.
The integrated flux density of NVSS sources is extracted from the Unified Radio Catalog
\footnote{\url{http://www.aoc.nrao.edu/~akimball/radiocat.shtml}},
which is a combined radio objects catalog with the correction of the flux and position.

The \fathomer and \crafts survey fields are both fully covered by the NVSS field. 
Both the \fathomer and \crafts surveys adopt the drift scan observation mode, which is scanning
at a constant declination within an R.A. range. 
During the drift scans, 
the rotation angle of the 19-feed array is fixed at $23.4^\circ$
to obtain the optimized field coverage \citep{2018IMMag..19..112L}.
Therefore, the same sources might be observed multiple times by different feeds.
In addition, we also selected the sources with their declination within $10'$ to the scanning declination 
to recover a full pattern beyond the size of the main beam.
We choose a total area of $\sim 138 \, {\rm \deg}^2$ and $\sim 385\, {\rm \deg^2}$
NVSS field for \fathomer and \crafts, respectively. 
The total number of sources within the survey areas are $7 \ 463$ and $21 \ 015$ 
for \fathomer and \crafts, respectively.
However, not all the sources are used in the beam pattern construction.
Some of the sources are dropped because they are either poorly observed or highly contaminated by 
RFI or nearby sources. 
The detailed source selection criteria are presented in \refsc{sec:sourceselection}.

We select a frequency range between $1375\,{\rm MHz}$--$1425\,{\rm MHz}$
which is relatively RFI-free and central at the NVSS observation frequency. 
We also assume a smooth power-law spectrum for the sources and adopt the 
NVSS median spectral index of $0.7$ \citep{1998AJ....115.1693C} to 
calculate the flux density at each frequency. 
The NVSS sources are assumed to be unresolved point sources and their flux densities
are then converted to brightness temperature assuming 
a constant gain factor, i.e. DPFU (Degree Per Flux Unit) 
$A_{\rm eff}/2k_{\rm B} = 25.6\,{\rm K}\, {\rm Jy}^{-1}$ \citep{2020RAA....20...64J}.

\section{Method} \label{sec:method}

\subsection{Beam stacking}

The time-ordered data is expressed as the combination of beam-convolved sky and noise,
\begin{align}
\bm d = \bm P \bm M \bm b + \bm n,
\end{align}
where $\bm d$ and $\bm n$ are the time-ordered data vector and the noise vector, 
with the vector element representing the flux measurements and the corresponding 
noise at each time stamp, respectively.
The multiplication of $\bm M \bm b$ represents the convolution of the sky model and 
the beam kernel.
$\bm b = \{\bm B_{1}^{\rm T}, \bm B_{2}^{\rm T}, \cdots, \bm B_{q}^{\rm T} \}^{\rm T}$ 
is a column vector by stacking the columns of beam matrix $\bm B$ with size $(q, q)$, where
$\bm B_{i}$ represents the $i$-th column of the beam matrix.
Matrix $\bm M$ is a special matrix of size $(p, q^2)$, where $p$ is the number of pixels
within the survey area. The matrix $\bm M$ captures the convolutional spatial relationships 
between the pixels of the sky and the beam kernel matrix $\bm B$, i.e. each row of the matrix
$\bm M$ is contracted by stacking the pixels covered by the beam kernel at each pointing direction. 
The matrix $\bm P$ is the projection matrix that maps the beam-convolved sky pixels to the time stamps.  

To get the estimation of the beam kernel $\bm b$ in the presence of noise, 
the $\chi^2$ is expressed as,
\begin{align}
\chi^2 = \left(\bm d - \bm{PMb}\right)^{\rm T} C_{\rm N}^{-1} \left(\bm d - \bm{PMb}\right),
\end{align}
where $C_{\rm N} = \langle \bm n\bm n^{\rm T} \rangle$ is the noise covariance matrix. 
As shown in previous work \citep[i.e.][]{2023ApJ...954..139L}, the FAST drift scan observation 
exhibit stable noise level and the correlated noise is eliminated with the 
temporal gain calibration. To simplify the estimation, the noise covariance matrix
is assumed to be an identity matrix. 
By minimizing $\chi^2$, the beam kernel estimator is expressed as,
\begin{align}
\hat{\bm b} = \left( \left(\bm{PM}\right)^{\rm T} C_{\rm N}^{-1} \left(\bm{PM}\right) \right)^{-1}
\left(\bm{PM}\right)^{\rm T}C_{\rm N}^{-1} \bm d.
\end{align}
The beam matrix is then expressed as, 
\begin{align}\label{eq:bm}
\hat{\bm B} = \{\hat{b}^{\rm T}_{1,\cdots,q}, 
                \hat{b}^{\rm T}_{q+1,\cdots,2q}, 
                \cdots, 
                \hat{b}^{\rm T}_{q(q-1)+1,\cdots,q^2}, 
                \},
\end{align}
where $\hat{b}_{i,\cdots,j}$ represents a vector composed with
the $i$-th to $j$-th elements of $\hat{\bm b}$.
We use a beam matrix with the shape of $100 \times 100$ and resolution of $12''$,
covering $20'$ in both dimensions, which is large enough to include the first few side lobes. 

We adopt the time-ordered data (TOD) pre-processed in the previous work of
\citet{2023ApJ...954..139L} and \citet{2024arXiv241208173Y} for the 
FATHOMER and CRAFTS data, respectively. 
The data process is performed on the two linear polarization auto-power, 
referred to as $d_{\rm XX}$ and $d_{\rm YY}$, 
which is then combined to form the Stokes I components,
\begin{align}\label{eq:I}
d_{\rm I} = (d_{\rm XX} + d_{\rm YY})/2.
\end{align}
It is worth noticing that, with the TOD pre-processing, the data is pre-whitened by 
subtracting the frequency-averaged temporal baseline.

\subsection{The sky model}

The sky model is constructed with the continuum flux density of sources from the NVSS catalog. 
Because the FAST beam size is $\sim 3'$, all the NVSS sources are assumed to be unresolved
and the sky model is expressed as,
\begin{align}
m_p = \sum_{i \in \mathcal{I}} S_i, \quad \mathcal{I} = 
\{ i \mid |\alpha_i - \alpha_p| < \frac{\delta}{2}, |\beta_i - \beta_p| < \frac{\delta}{2} \}
\end{align}
where $(\alpha_i,\beta_i)$ represents the coordinate of the $i$-th sources with flux density of $S_i$
and $m_p$ is the total flux density of the pixel at $(\alpha_{p}, \beta_{p})$
with pixel size of $\delta$.
In this work, the pixel size of the map is set to match the beam model kernel, i.e. $\delta = 12''$.
The pixel size in this work is sufficiently small relative to the FAST beam width of $\sim 3'$, 
ensuring minimal position accuracy loss due to pixelation.

\subsection{Source selection criteria}\label{sec:sourceselection}

\begin{figure*}
\centering
\includegraphics[width=\columnwidth ]{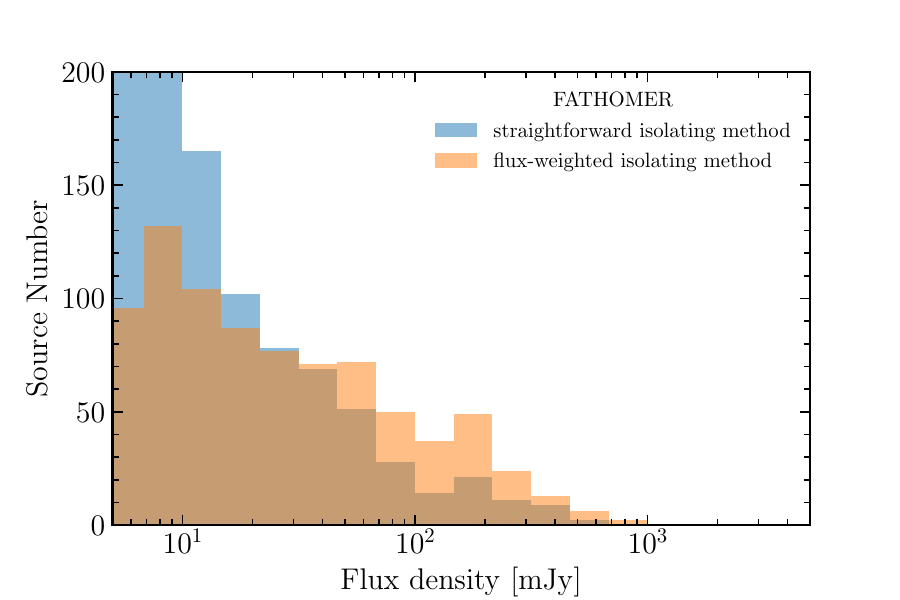}
\includegraphics[width=\columnwidth ]{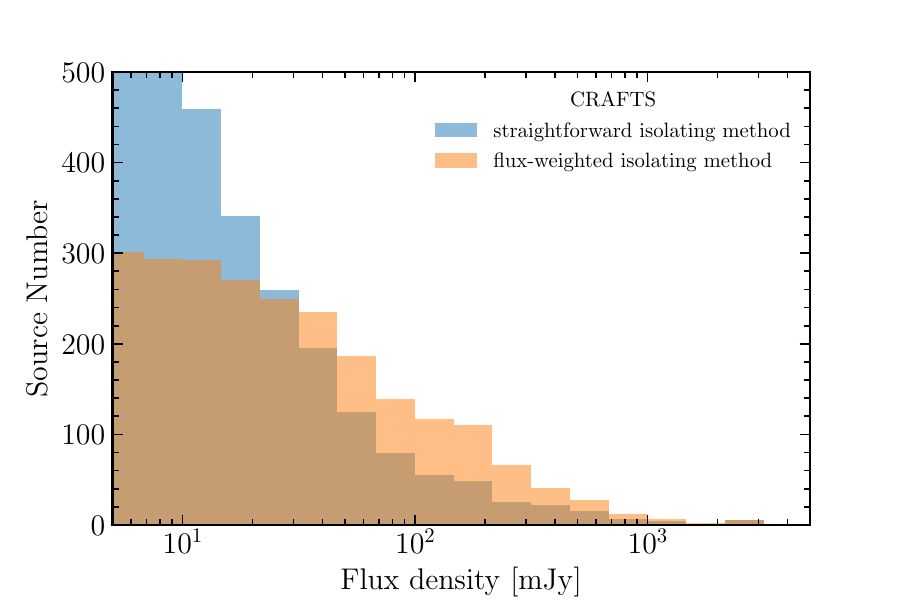}
\caption{
The histogram distribution of source number counts for the two source isolating methods. 
The straightforward isolating method result is shown in blue and the flux-weighted isolating method result is in orange.
The left panel shows the results of the \fathomer dataset and the right panel shows the
results of the \crafts dataset, respectively.
}\label{fig:Flux_count}
\end{figure*}

The matrix $\bm P$ represents the projection between the maps and the time-ordered domains. 
Thus, it has the shape of $t \times p$, where $p$ is the number of pixels of the sky model and
$t$ is the number of time stamps. 
Utilizing the complete time-ordered observational data yields a large $\bm P$ matrix, 
rendering the subsequent matrix operation both time- and memory-intensive.
On the other hand, not all the observational data are useful for beam measurements. 
For instance, the RFI-contaminated data are removed at the first stage of data selection.
Although we chose a frequency band that is relatively RFI-free,
there still are time stamps that are badly contaminated by transient objects, e.g. 
the navigation satellites. 
In addition, calibrators with other sources nearby are not ideal for characterizing the beam pattern.
Thus, we only chose the time-ordered data
with bright isolated sources for the following beam stacking analysis, which significantly reduces the
amount of observation data.

A straightforward isolating method is to remove the sources with any other sources within the 
radius of $5'$. We apply the straightforward isolating method to the initial NVSS catalog 
before applying the $20\,{\rm mJy}$ flux threshold. 
Such a straightforward isolating method significantly reduces the amount of 
observation data. However, it indiscriminately removes both the bright and dim sources.

To keep as many bright sources as possible, we consider a flux-weighted isolating method.
We apply a point spread function (PSF) to each of the sources and calculate
the ratio of the source spread flux to the target central sources,
\begin{align}
\epsilon_{ij} = S_i {\rm PSF}(\delta \alpha \cos(\beta), \delta \beta) / S_j,
\end{align}
where $S_i$ is the flux density of the neighboring source and $S_j$ is the flux density of the central source,
$(\delta \alpha \cos(\beta), \delta\beta)$ are the separation angle between the 
two sources in the R.A. and declination directions. The PSF is assumed to be a Gaussian kernel with 
$\sigma_{\rm PSF} = \theta_{\rm FWHM}/2\sqrt{2\ln 2}$, where $\theta_{\rm FWHM} = 2.9'$ is the Full Width at Half Maximum of FAST prior beam.
We set a threshold of $\epsilon = 0.05$, i.e. the sources are removed if there is more than one nearby source's point-spread flux density over $5\%$ of the target source flux density.


Using the straightforward isolation method, we selected a total of $7\ 159$ sources, 
e.g., $1\ 798$ from the \fathomer field and $5\ 361$ from the \crafts field. 
When applying the flux-weighted isolation method, the number of selected sources decreased 
to $4\ 202$, with $1\ 087$ in the \fathomer field and $3\ 115$ in the \crafts field.
Figure~\ref{fig:Flux_count} presents the flux density distribution of sources 
identified using the two isolation methods. Although the flux-weighted isolation 
method reduces the total number of sources, it retains more bright sources 
than the straightforward method, enhancing the quality of beam pattern reconstruction. 
Applying a final flux limit of $20\,{\rm mJy}$, the flux-weighted isolation method selects
$1\,668$ sources, which is $456$ more than the straightforward method.

Before proceeding to the beam stacking, 
we further truncate the time-ordered data and keep a segment of data with 
only $50$ time stamps before and after the transition time, which is defined 
as the time stamps when the target sources are mostly close to the beam center.
It should be noted that each source can be observed multiple times by 
different beam at different times.
Considering the observation duplication, there are $\sim 960$ and $\sim 1\,200$ 
data segments per beam in the \fathomer and \crafts field, respectively, 
finally used for the beam pattern construction.

\subsection{Zernike polynomial beam model}

We adopt the Zernike polynomial (ZP) modes as the analytic basis and 
decompose the 2-dimensional beam pattern against these modes.
The ZP modes are a sequence of polynomials that are continuous and orthogonal
over a unit circle \citep{1934MNRAS..94..377Z}. 
Initially, they were utilized as a mathematical representation of 
optical wavefronts traversing imaging components \citep{2011JMOp...58..545L}, 
and then employed to characterize the primary beam patterns for radio telescopes
\citep{2021MNRAS.502.2970A}. 
The ZP modes of order $n$ and angular frequency $m$ take the form of, 
\begin{align}
Z^{m}_{n}(\rho, \phi) = R^{m}_n(\rho) e^{i m \phi},
\end{align}
where $(\rho,\phi)$ are the coordinates of the polar frame with origin at the beam center,
$n$ and $m$ are non-negative integers with $n>|m|$ and 
$n-|m|$ is always even. The radial polynomials $R^{\pm m}_n(\rho)$ is expressed as,
\begin{align}
R^{\pm m}_n(\rho) = \sum_{k=0}^{\frac{n-m}{2}} \frac{(-1)^k (n-k)!}{k! (\frac{n+m}2 - k)! 
(\frac{n-m}2 - k)!} \rho^{n-2k}.
\end{align}

The beam pattern is constructed as a linear combination of these ZP modes,
\begin{align}
\hat{{\bm B}} = \sum_{i}^{N_{Z}} a_i {\bm z}_i, 
\end{align}
where $\hat{\bm B}$ is the beam pattern constructed via stacking analysis, i.e. \refeq{eq:bm},  
$i$ denotes the Noll indices \citep{1976JOSA...66..207N} of the ZP mode $Z^{m}_{n}$,
$N_Z$ is the number of ZP modes used for beam model construction,
${\bm z}_i$ denotes the $i$-th ZP modes, i.e. ${\bm Z} = \{{\bm z}_i\} = \{Z^{m}_{n}(\rho, \phi)\}$,
and $a_i$ represents the corresponding coefficients,
\begin{align}
{\bm A} = \{a_i\} = \left({\bm Z}^{\rm T} {\bm Z}\right)^{-1} {\bm Z} {\bm B}.
\end{align}


The ZP modes with higher orders represent the extremely small-scale structures, 
which are noise-dominated. 
We use the ZP modes up to the $23$rd order, which contains three hundred coefficients. 
In addition, the coefficients of the ZP modes within the first $23$ orders vary significantly.
In order to improve the signal-to-noise ratio of the beam pattern model, we drop the 
noise-dominated ZP modes. The detailed noise filtering is presented in \refsc{sec: 19beams}.

%

\section{Results and Discussion} \label{sec:results}

\begin{figure}
\includegraphics[width=0.85\columnwidth]{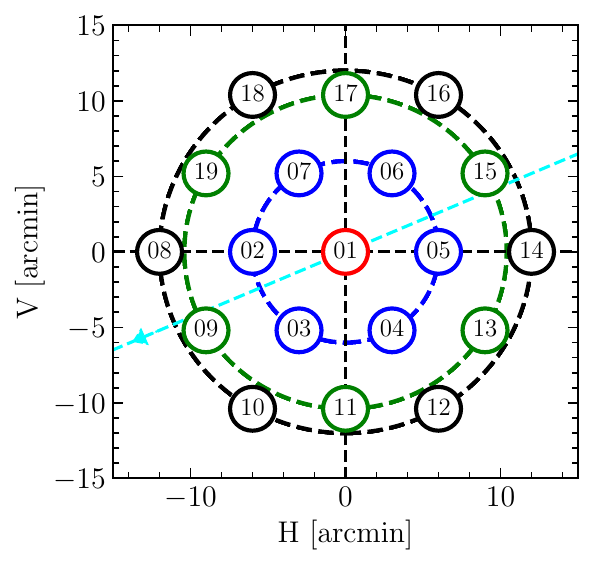}
\caption{
    The positions of the feeds in the frame of the FAST L-band 19-feed array (19FA). 
    The circles are the FAST beams with the beam size $2.9'$. 
    The center, inner-circle, middle-circle, and outer-circle beams are shown in 
    red, blue, green, and black, respectively.
    The cyan dashed line with an arrow indicates the drift scanning direction along R.A..
}
\label{fig: BeamPos}
\end{figure}

The positions of the 19 feeds within the FAST L-band 19-feed array (19FA) frame
are shown in \reffg{fig: BeamPos}. Here, the $H$-axis and $V$-axis correspond 
to the horizontal and vertical feed axes.
We group the 19 feeds into four subsets according to their separation distance 
to the feed array center, i.e. as shown in \reffg{fig: BeamPos},
the center beam (red), inner-circle beams (blue), middle-circle beams (green), 
and the outer-circle beams (black). 
During the drift scan observation, 
the rotation angle of the feed array was fixed at $23.4^\circ$.
The cyan dashed line with an arrow indicates the drift scanning direction,
which is aligned with the R.A. direction in the observational frame.

\subsection{The stacking-based beam pattern}\label{sec:stackingbeam}

\begin{figure}
\includegraphics[width=\columnwidth]{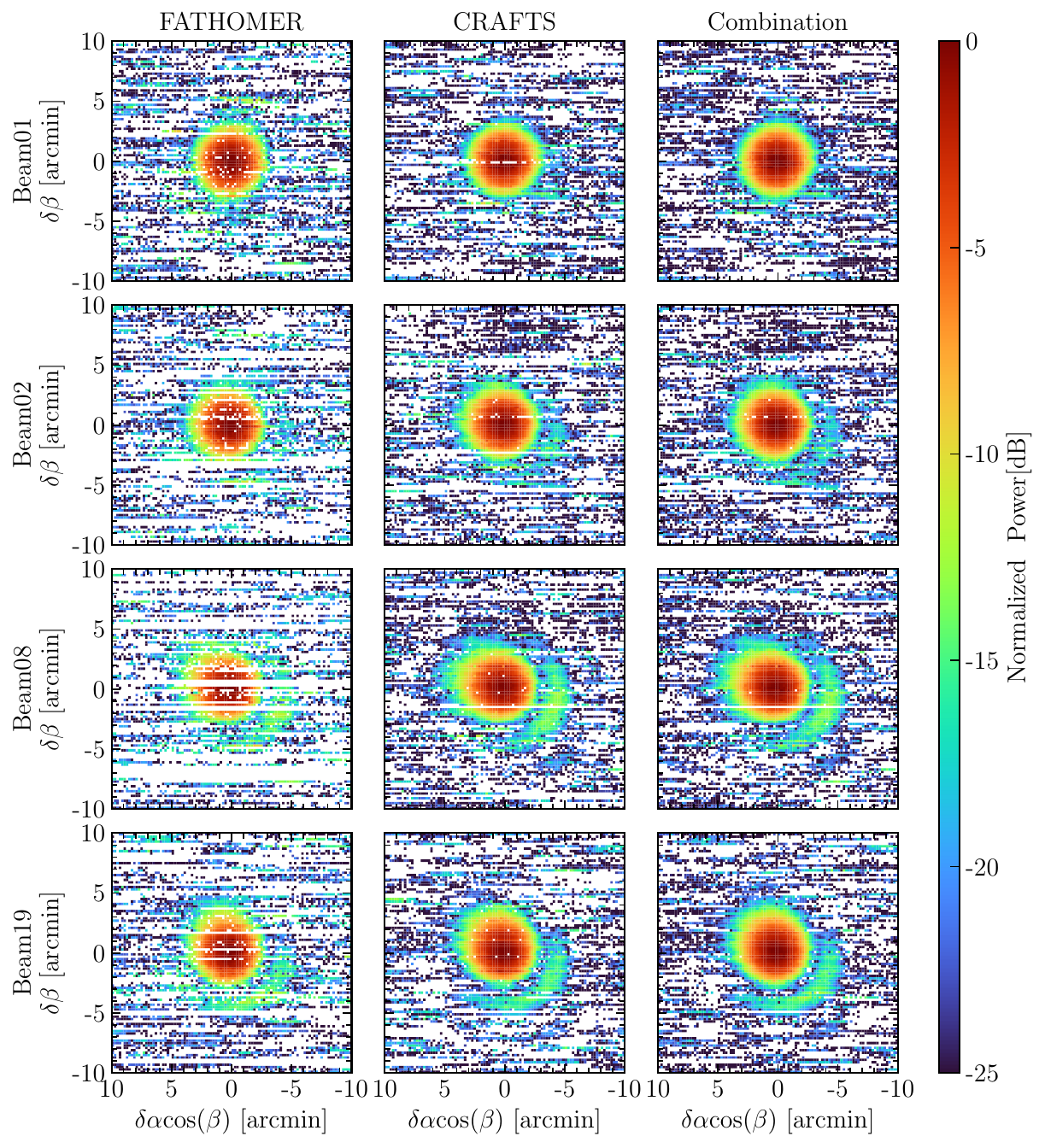}
\caption{
The stacked beam patterns for four different beams (Beam01, Beam02, Beam08, and Beam19) were obtained from the data between 1375-1425MHz. The three columns represent the data from three different sources: \fathomer (first column), CRAFTS (second column), and a combination of both (third column). Each row corresponds to a unique beam, with each beam having a distinct distance from the central beam center.
}
\label{fig: combination}
\end{figure}

\begin{figure}
\includegraphics[width=\columnwidth]{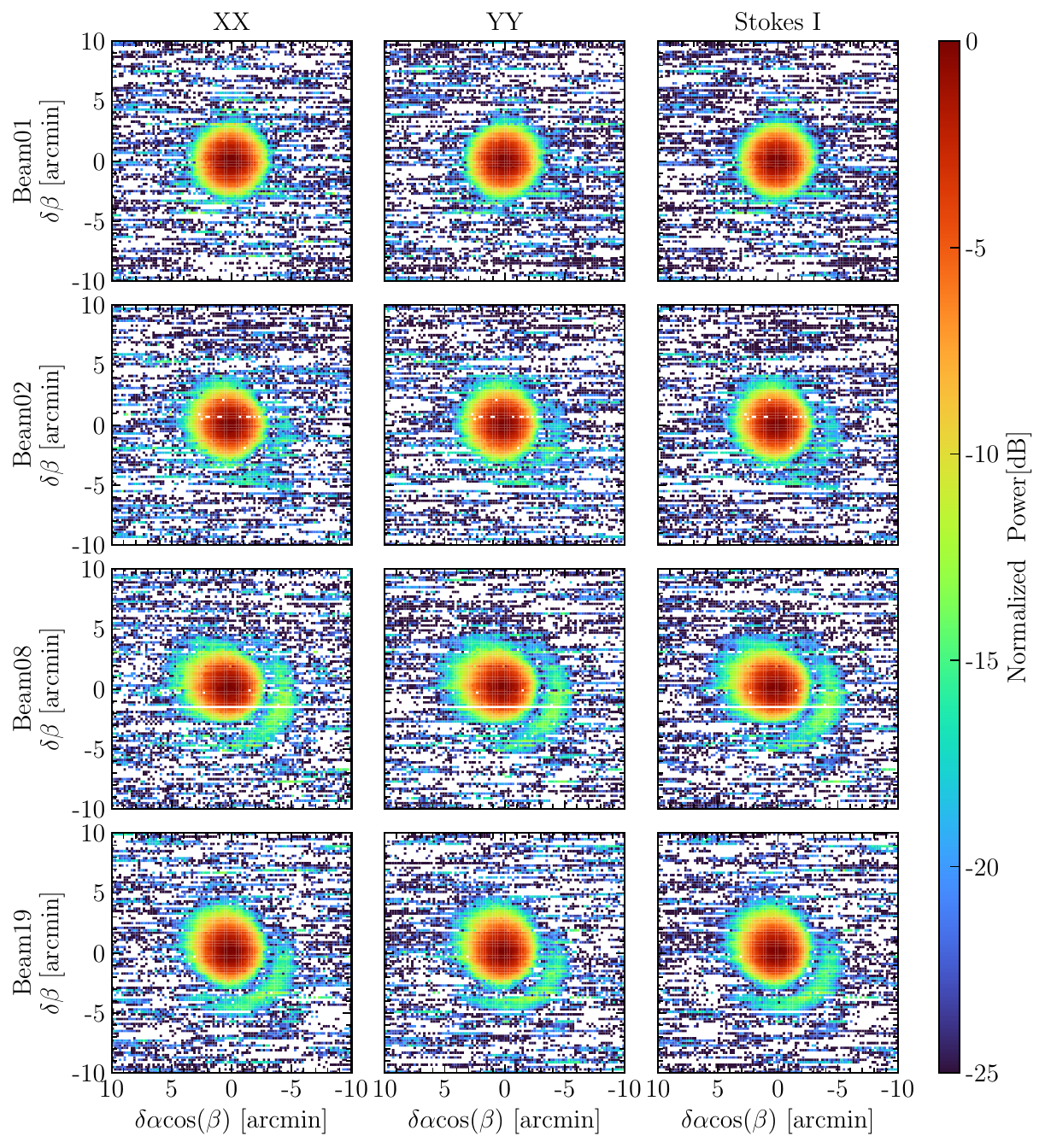}
\caption{
The stacked beam patterns for four different beams (Beam01, Beam02, Beam08, and Beam19) were obtained from the data between 1375-1425MHz. The three columns represent the polarization results for XX (vertical), YY (horizontal), and Stokes I (total intensity), respectively. Each row corresponds to a unique beam, with each beam having a distinct distance from the central beam center. 
}
\label{fig: polar}
\end{figure}

To investigate the consistency of the stacking-based beam pattern reconstruction method, 
we select observations with similar declination coverage from \fathomer and \crafts.
We then perform the stacking analysis separately on each dataset as well as on the combined Stokes I dataset
via \refeq{eq:I} to increase the signal-to-noise ratio. 
The reconstructed beam patterns are shown in \reffg{fig: combination}, where the results of \fathomer, \crafts, and their combined dataset are shown in 
the left, middle, and right columns, respectively. 
From the top to the bottom rows, the results are Beam01, Beam02, Beam19, and Beam08, 
which are examples of beams located in the center, inner circle, middle circle, and
the outer circle of the FAST L-band feed array, respectively. 
All the beam patterns are shown in the observation frame, 
where the horizontal and vertical axes are aligned with the R.A. and declination direction. 

The shape of the beam pattern can be constructed by stacking either with 
the \fathomer or \crafts dataset.
The circular patterns at the center are identified as the main lobe, 
while the asymmetric patterns offset from the center are referred to as side lobes. 
A detailed definition of the main lobe is provided in \refsc{sc:beameff}.
The side lobes of Beam02, Beam19, and Beam08 are 
visible with each dataset. However, significant data deficiency, which is due to 
either lack of observation or RFI flagging, is also visible, 
even within the main lobe.
Such data deficiencies are substantially eliminated by combining the two datasets,
as shown in the right columns, where 
some blank pixels within the main lobe are filled by the complementary dataset.
The remaining blank pixels, particularly those outside the main beam with negative power values, 
are likely due to observational noise, as TOD pre-whitening involves subtracting the temporal baseline.

The shapes of the main beam and side lobes show coma distortions across 
different beam positions. Beam01 displays a symmetric circular main beam profile, 
while Beam02, Beam08, and Beam19 exhibit more pronounced distortions, each showing an 
asymmetry with one side compressed and the other elongated. 
Notably, the side lobe shapes are the most striking feature among the four beams. 
Beam01’s side lobe is suppressed, whereas those of Beam02, Beam08, and Beam19 are visible. 
The side lobes of Beam08 and Beam19, in particular, are especially pronounced, 
with clearly discernible contours.

The direction of asymmetry distortion is directly related to the beam's position 
with respect to the center of the 19FA frame. 
The greater the distance of the beam from the center of the feed array, the more pronounced the asymmetry of the beam becomes.



In addition to the polarization-combined beam stacking, we also investigate the 
beam patterns for the XX and YY polarization, individually.
As only a few of the selected NVSS sources are significantly polarized, 
we ignore the polarization fraction and assume equal flux density for each polarization. 
The results are shown in \reffg{fig: polar}, where the left and middle columns show the
beam pattern for XX and YY polarizations.
The polarization-combined results are shown as a reference in the right column.
Similar to \reffg{fig: combination}, 
we also inspect the beam patterns of Beam01, Beam02, Beam08, and Beam19, as the
examples for the center, inner-circle, outer-circle, and middle-circle beams.
All the beam patterns shown in \reffg{fig: polar} use the combined 
datasets of \fathomer and \crafts. 

We find that the beam features for both XX and YY polarizations are similar, 
and align with the polarization-combined beam patterns.
Given the current sensitivity limit, 
the differences between polarizations are negligible and we only focus on the 
polarization-combined beam pattern, i.e. the beam pattern of Stokes I, in the following 
analysis.


%

\subsection{The beam pattern model}

\begin{figure*}
\center
\includegraphics[width=\textwidth]{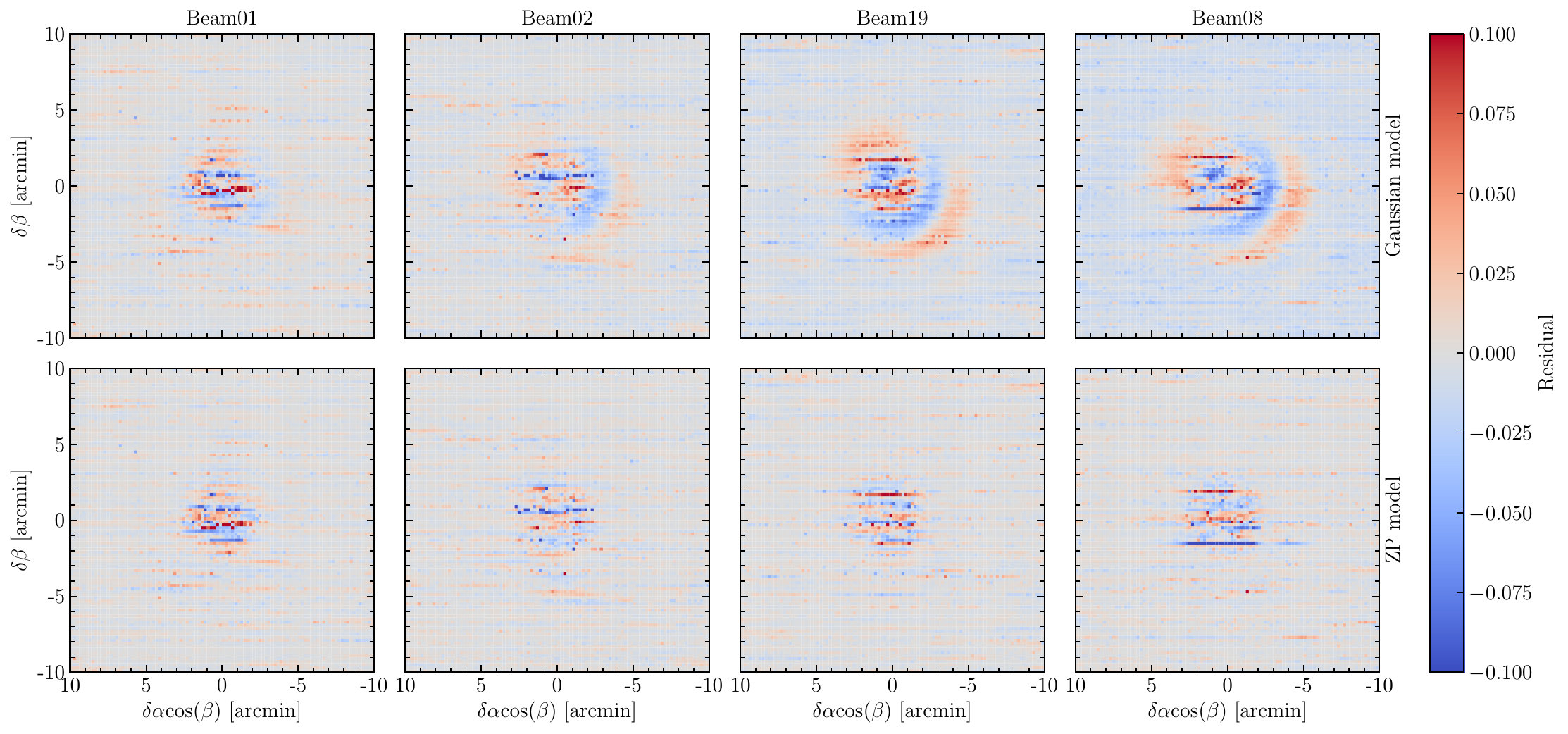}
\caption{
The residuals of the measured beam pattern and the fitting models. 
These residuals are divided by the main lobe peak value of fitting models.
The top panels show the residuals for the Gaussian model while the bottom panels show
the residuals for the ZP model.
From the left to the right columns, there are the residuals for Beam01, Beam02, Beam19,
and Beam08, which are examples of the subset of the center beam, inner-circle beams, 
middle-circle beams, and outer-circle beams. 
}
\label{fig:Residual}
\end{figure*}

Since the beam pattern shows significant asymmetry, we adopt the ZP modes as the 
analytical bases to model the beam pattern features. In particular, we use
the Python script of \texttt{zernike} \footnote{\url{https://pypi.org/project/zernike/}} 
\citep{2015JOSAA..32.1160A} to generate the ZP modes and decompose the beam patterns.

As a comparison, we also fit the beam patterns with a circular-symmetric Gaussian profile,
\begin{align}
B_{\rm G}(\rho, \phi) = \exp\left(-\frac{1}{2}\frac{\rho^2}{\sigma_{\rm G}^2}\right),
\end{align}
where $(\rho, \phi)$ are the coordinates of the beam-centered polar frame,
$\sigma_{\rm G}$ is a free parameter characterizing the beam size and related to
the full width at half maximum  $\theta_{\rm FWHM}$ of the beam pattern via
$\sigma_{\rm G} = \theta_{\rm FWHM}/\left(2\sqrt{2\ln 2}\right)$.

To quantify the reconstruction accuracy, we calculate the residual between the 
measured beam pattern $\hat{B}(\rho, \phi)$ and the beam model for both the ZP modes
and Gaussian profile,
\begin{align}
r(\rho, \phi) = \hat{B}(\rho, \phi) - B_{\{{\rm G,Z}\}}(\rho, \phi),
\end{align}
where $\{{\rm G, Z}\}$ denote the Gaussian and ZP model, respectively.
We show examples of the residuals for both the Gaussian model (top panels) and 
the ZP model (bottom panels) in \reffg{fig:Residual}. 
From the left to the right column, we show the residuals of Beam01, 
Beam02, Beam19, and Beam08, which are examples of the center beam, 
inner-circle beams, middle-circle beams, and outer-circle beams, respectively. 
It is clear that, for the center beam, the Gaussian model and ZP model have 
similar residual patterns. However, for the beams off the feed array center, 
there are significant residual patterns for the Gaussian model.

\begin{figure*}
\center
\includegraphics[width=0.95\textwidth]{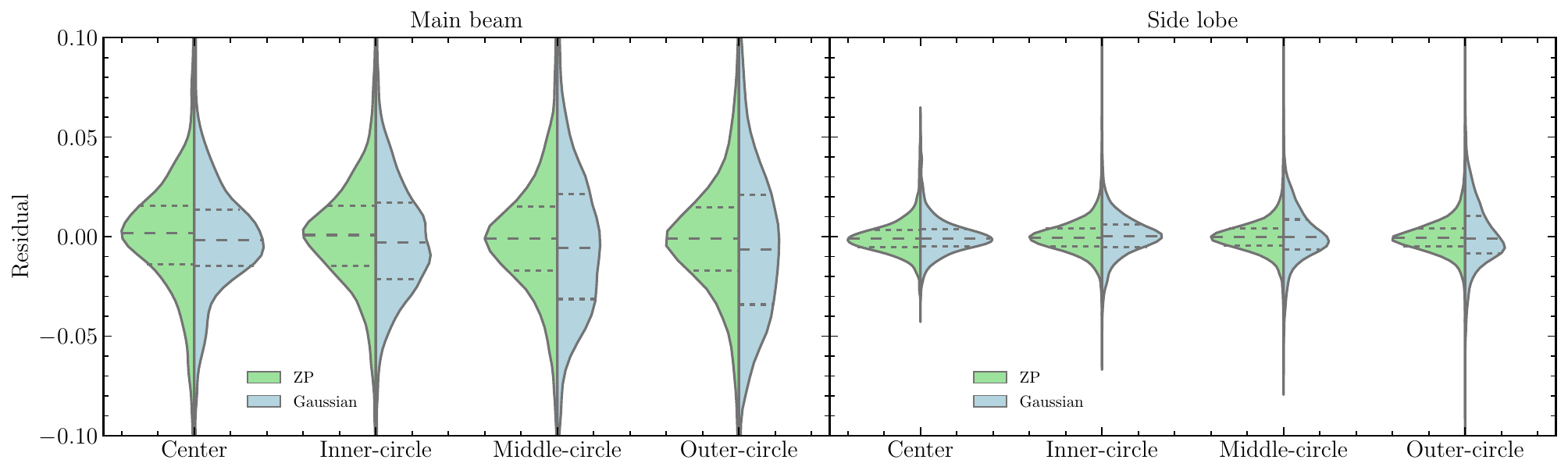}
\caption{
The residual histogram distribution of all the beams in the same beam subset, 
i.e. the center beam, inner-circle beams, middle-circle beams, and outer-circle beams.
The residuals for the ZP model are shown in green (left side of the violin plot) 
and residuals for the Gaussian model are shown in blue (right side of the violin plot).
The left panel shows the residuals within the main beam size, i.e. $\rho< \theta_{\rm FWHM}$;
while the right panel shows the residuals of the side lobe, i.e. $\theta_{\rm FWHM}\leqslant\rho<2\theta_{\rm FWHM}$.
}
\label{fig:ResDensity}
\end{figure*}

We inspect the residual for all the beams of the same subsets via the 
statistical histogram distributions with respect to the $\rho$ bins.
In addition, we characterize the residual distribution for the main beam
and the first side lobe, separately. According to the FAST beam measurements
in the literature, we set a prior of $\theta_{\rm FWHM}$, which corresponds to the radius of the first null,
for separating the main beam and side lobes,
i.e., for the main beam we use residuals within $\rho < \theta_{\rm FWHM}$ and for the side lobe 
we use $\theta_{\rm FWHM}\leqslant\rho<2\theta_{\rm FWHM}$. The residual statistical histogram as 
the function of radial distance $\rho$ for the main beam and first side lobe 
are shown with the violin plots in the left and right panel of \reffg{fig:ResDensity}, respectively. 
The residuals for the ZP model are shown in green (left side of the violin plot) and 
for Gaussian model are shown in blue (right side of the violin plot).
In each panel of \reffg{fig:ResDensity}, we show the residual histogram of the beams 
in the subset of the center beam, inner-circle beams, middle-circle beams, 
and outer-circle beams, individually. 

Generally, the residuals within the main beam have a broader histogram distribution 
than that in the side lobe, which is mainly due to the greater absolute values
of the main beam.






For the center beam, both models yield similar results, i.e., the residual distributions closely 
resembled the normal distribution. As long as the beam profile is circularly symmetric and 
the coma lobes are suppressed, the Gaussian profile is still a good approximation of the beam 
pattern shape.

In contrast, the residual distributions of the off-center beams, e.g., the inner-circle beams,
middle-circle beams, and outer-circle beams, reveal significant differences between the ZP model and 
the Gaussian model. Particularly, the Gaussian model has residual distributions
that deviate from the normal distribution, significantly, which indicates
the systematic differences between the Gaussian model and the beam pattern measurements. 
On the other hand, the ZP model residual behaves in the unique normal distribution for
all the beam subsets. The results show that the ZP model successfully characterizes the 
primary features of the beam pattern for both the symmetric and asymmetric cases,
as the ZP includes coma distortion terms.


\subsection{The ZP mode decomposition }

\begin{figure*}
\center
\includegraphics[width=0.9\textwidth]{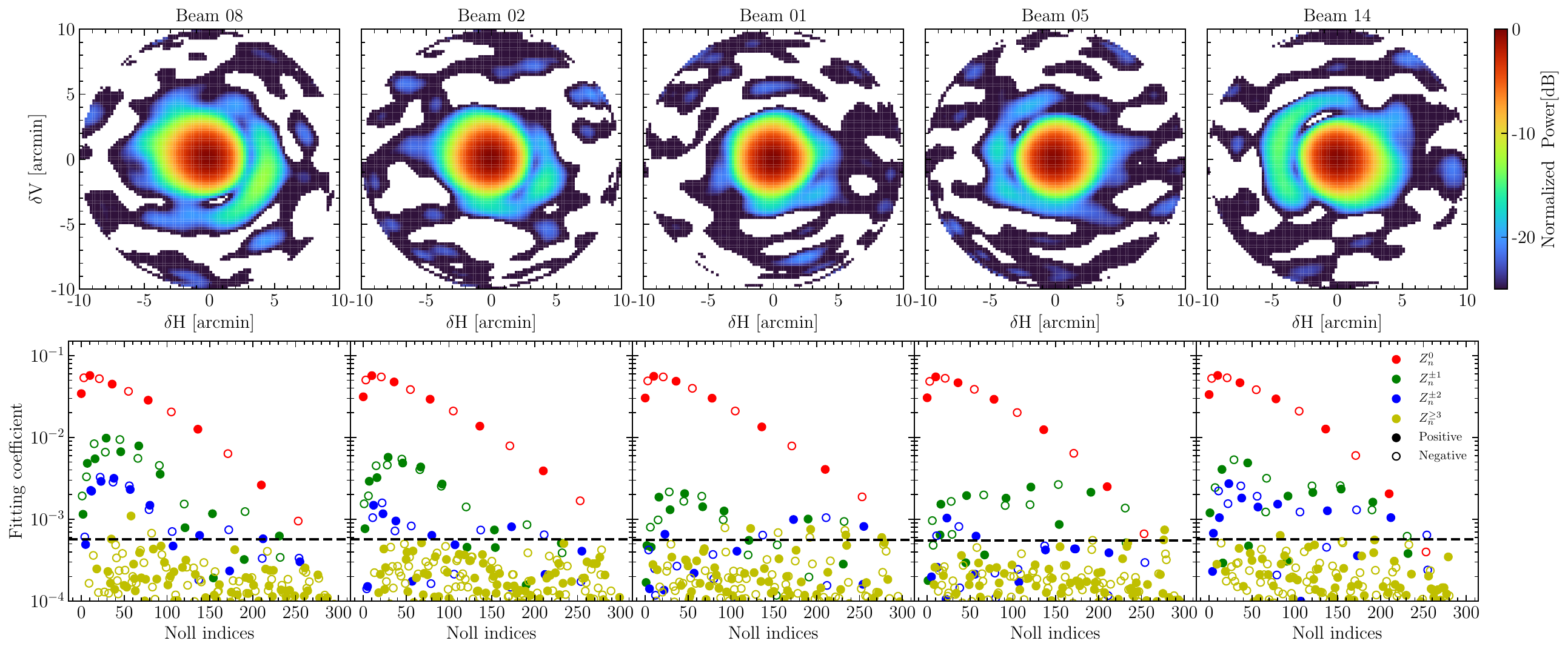}
\caption{
Top panels: the constructed beam pattern with the first 23 orders of ZP modes. 
All the beam patterns are rotated to the 19FA frame.
The $\delta H$- and $\delta V$-axes represent the separation angle 
in the $H-V$ axes with respect to the beam center.
Bottom panels: the corresponding ZP fitting coefficients shown as
the function of Noll indices.
The filled markers denote positive coefficients and the empty markers 
indicate negative coefficients. The colors mark different angular frequencies $m$.
The horizontal dashed line indicates $1\%$ of the maximum fitting coefficient.}
\label{fig: beamfit}
\end{figure*}

\begin{figure*}
\center
\includegraphics[width=0.9\textwidth]{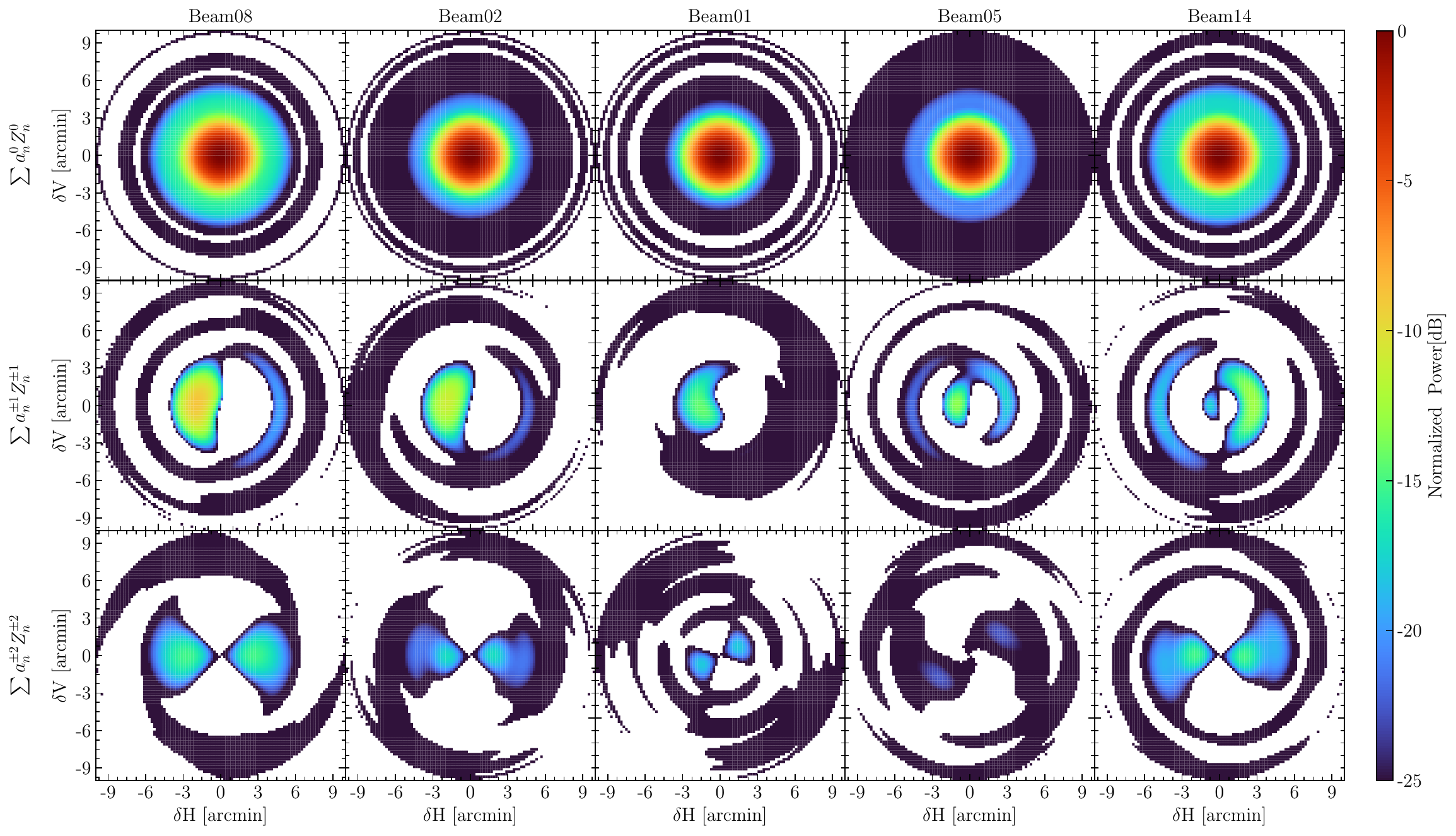}
\caption{The beam pattern constructed with ZP mode of different angular frequency $m$,
i.e. the beam pattern constructed with ZP modes of $m=0$, $m=\pm 1$, and 
$m = \pm 2$ are shown in the top, middle, and bottom rows, respectively.
}
\label{fig: Zde}
\end{figure*}

The top panels of \reffg{fig: beamfit} display the ZP-model reconstructed beam patterns, 
where we display instances of the $5$ beams in a row aligned with the 
$H$-axis of the 19FA frame.
We rotate the beam pattern from the observation frame to the 19FA frame.
The center beam shows a circularly symmetric main beam profile and suppressed side lobes, 
while the off-center beams show significant asymmetry and pronounced side lobes.
The side lobes of the off-center beams are distinctly pronounced toward the 
center of the 19FA frame, and the beam distortion aligns with the radial direction of the frame.

The bottom panels display the fitting coefficients with respect to the 
Noll indices, up to $300$, which is corresponding to the first $23$ orders of the ZP modes.
The ZP modes with lower Noll indices correspond to the smaller variation structures. 
By analyzing the coefficients of these polynomials, we can gain insights into 
the beam pattern structures and their relationship to the side lobe intensities.
Specifically, the variation in the coefficients of the polynomials with circular frequency $m$ 
across different beams suggests that the beam pattern structure 
is not uniform and is influenced by various factors. 

The fitting coefficients of the ZP modes with angular frequencies $m = 0$, $\pm 1$, $\pm 2$,
and $|m| \geqslant 3$ are plotted in red, green, blue, and yellow, respectively. 
The solid and empty markers distinguish the positive and negative values, respectively.
As shown in the plot, the ZP modes with an angular frequency $m=0$ exhibit relatively large 
coefficients for all the beams, which demonstrates the beam patterns are primarily 
characterized by such ZP modes. We display the beam patterns constructed with only the
ZP modes of $m=0$ in the top rows of \reffg{fig: Zde}.
The ZP modes of $m=0$ characterize the circularly symmetric features of the beam patterns. 

The circular asymmetric features are characterized by the ZP modes with $m \neq 0$.
As shown in \reffg{fig: beamfit}, the corresponding fitting coefficients of such modes,
i.e., $Z^{\pm 1}_n$, $Z^{\pm 2}_n$, and $Z^{|m| \geqslant 3}_n$,
are significantly lower than the coefficients of $Z^{0}_n$. 
Moreover, the center beam 
exhibits even lower fitting coefficients for the ZP modes with $m \neq 0$ than the off-center beams.
The beam patterns constructed with only the ZP modes of $m=\pm 1$ and $m=\pm 2$ are
shown in the middle and bottom rows of \reffg{fig: Zde}.
The asymmetric features primarily characterize the distortion of the main beam
and the side lobes. 
Notably, as shown in the middle panels of such two rows,
there are weak asymmetry features in the center beams, which indicates a slight distortion 
of the center beam.
The origin of such center beam distortion is worth further investigating in the future with more
observation data.

\subsection{The noise-filtered beam pattern}\label{sec: 19beams}

\begin{figure*}
\center
\includegraphics[width=0.9\textwidth]{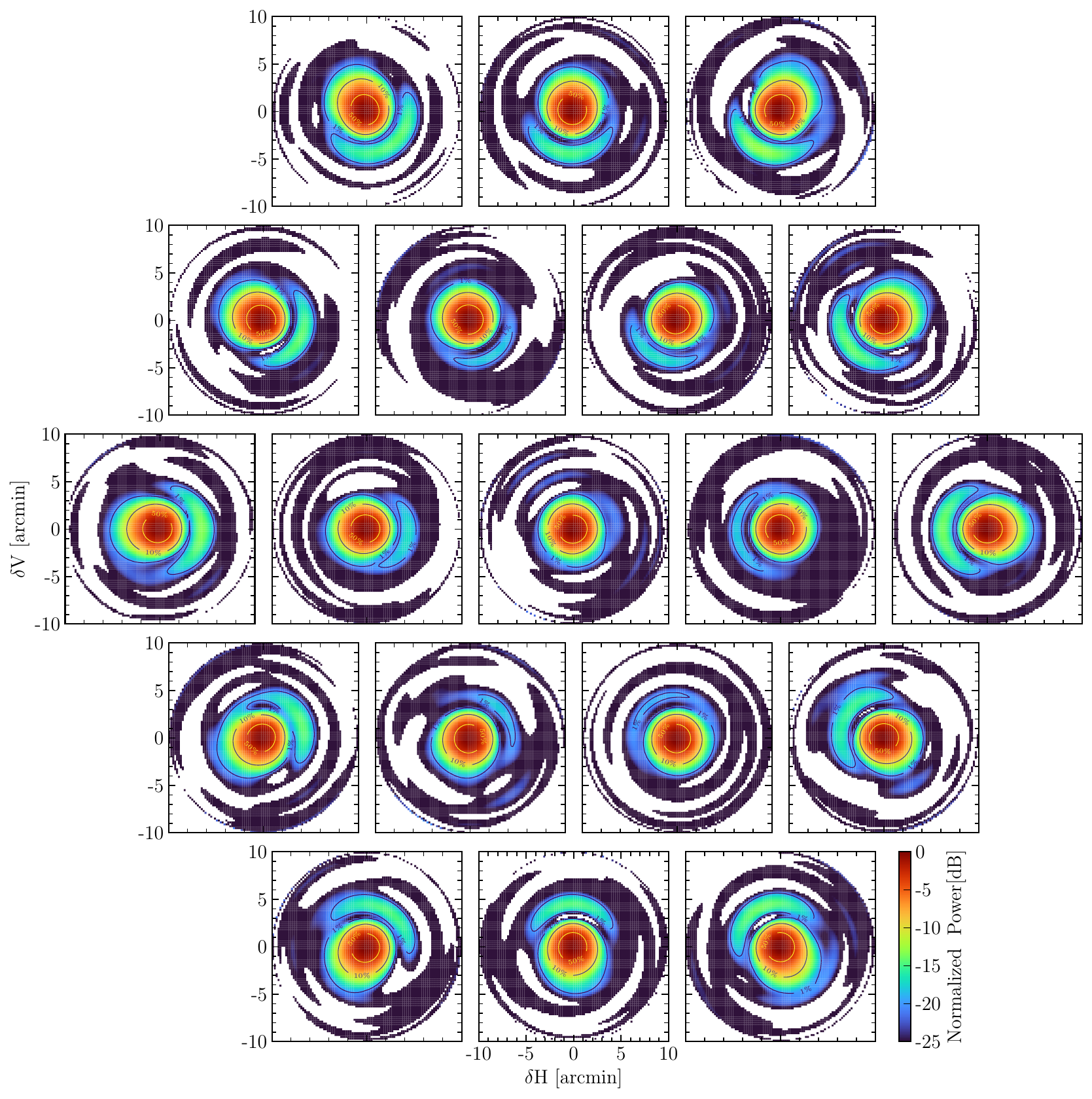}
\caption{The noise-filtered beam pattern of all the 19 beams of the FAST L-band feed array.
The beam patterns are rotated to align with the 19FA frame.
The positions of the beams correspond to their projected pointing direction in the sky.
The yellow, green, and blue contours represent the $50\%$, $10\%$, and $1\%$ of maximum 
response value at the beam center.}
\label{fig: contour}
\end{figure*}

The fitting coefficients of ZP modes with $|m| \geqslant 3$, as shown with the 
yellow markers in \reffg{fig: beamfit}, are mostly lower than $1\%$ of the 
maximum fitting coefficients.
ZP modes with $|m| \geqslant 3$ generally characterize highly oscillated structures. 
Although such structures are tiny compared to the main beam profile, it is
still important to be fully characterized for future \hiim experiments. 
However, given current measurement uncertainties, such weak structures are mostly dominated 
by observational noise. The equipment's thermal noise is one of the origins of 
measurement uncertainty. On the other hand, the flux density variation of the NVSS 
sources potentially impacts the measurement. Both the thermal noise and flux density 
variation are systematic and can be eliminated via a large amount of observation. 
In this work, we drop the ZP modes with fitting coefficients less than $1\%$ of the 
maximum fitting coefficients. 
Eventually, the number of ZP modes adopted varies across different beams.
The central beam has $31$ ZP modes, which is the minimal number of ZP modes.
The ZP mode numbers increase with the beam getting further from the feed array center, 
e.g., the inner-circle beams have $\sim 33$ ZP modes on average, while the middle- and outer-circle
beams have $\sim 44$ on average. 
The noise-filtered beam patterns for all the 19 feeds are shown in \reffg{fig: contour},
where the relative positions of the panels correspond to the feed pointing direction in the sky. 
All the beam patterns are rotated to the feed coordinates.
The yellow, green, and blue contours in each panel indicate the $50\%$, $10\%$,
and $1\%$ of the maximum, respectively.
As discussed in \refsc{sec:stackingbeam}, we focus on the Stokes I component and 
all the beam patterns are averaged across the frequency range of 
$1375\,{\rm MHz}$ -- $1425\,{\rm MHz}$.





The yellow contours in each panel indicate the beam pattern at $50\%$ of their maximum,
which is commonly defined as the beam size of full width at half maximum.
The beam profiles within the $50\%$ contour show unique circular symmetric and
the differences between the 19 beams are negligible.

The beam features between the $1\%$ and the $10\%$ contours are identified as the side lobes.
Significantly, the side lobe profiles vary across the 19 beams.
After filtering out the noise components, the asymmetric side lobe profiles are 
clearer than those shown in \refsc{sec:stackingbeam}. 
The coma lobes are getting more pronounced with the larger separation distance 
between the beam and the feed array center.
In addition, the direction of the side lobe distortion relates to the 
feed array’s relative position. Specifically, the coma lobes become more pronounced 
on the side of the beam closer to the center of the feed array along the radial direction.
Although the circular symmetry of the off-center beam is disrupted, 
the beam pattern still retains axial symmetry, with its axis of symmetry 
aligned with the radial direction of the feed array.

The ZP mode coefficients of the noise-filtered beam model can be
accessed from the GitHub repository {\tt Fast19FABM}
\footnote{\url{https://github.com/Nyarlth/Fast19FABM.}}.

\subsection{The side lobes}

\begin{figure}

	\includegraphics[width=\columnwidth]{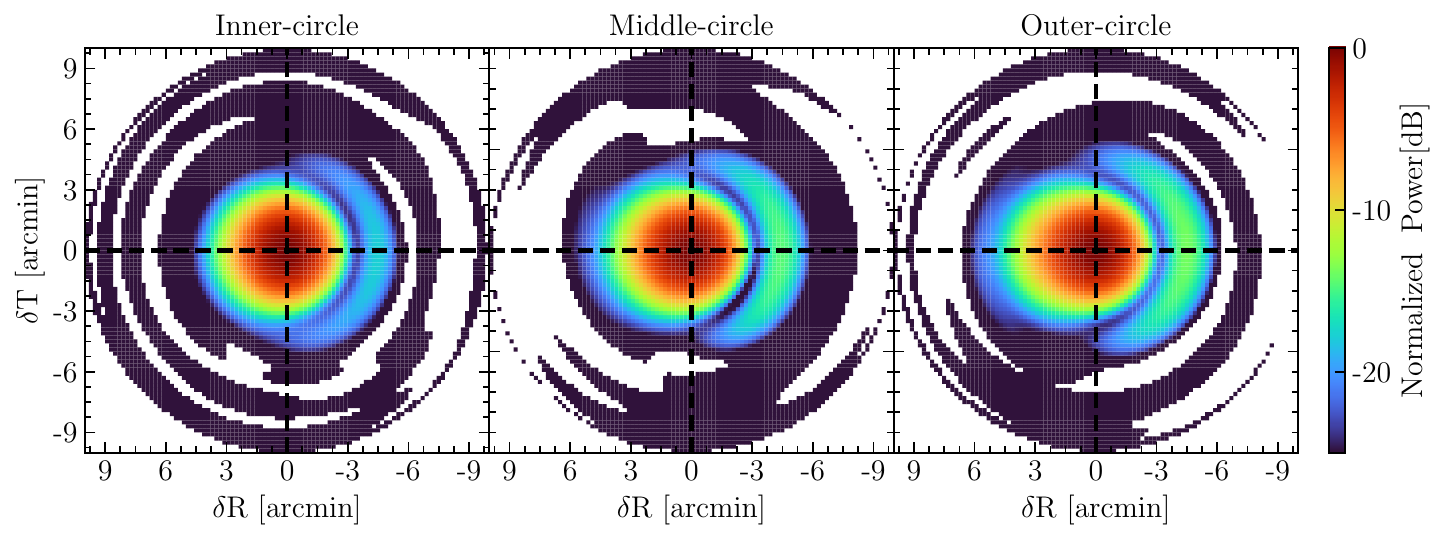}
    \caption{The beam pattern stacked with the beam in the subset of inner-circle (left panel),
    middle-circle (middle panel), and outer-circle (right panel) beams.
    The horizontal and vertical black dashed lines indicate the radial and tangential direction
    of the 19FA frame, respectively.
    }
    \label{fig:groupstack}
\end{figure}

\begin{figure}

	\includegraphics[width=\columnwidth]{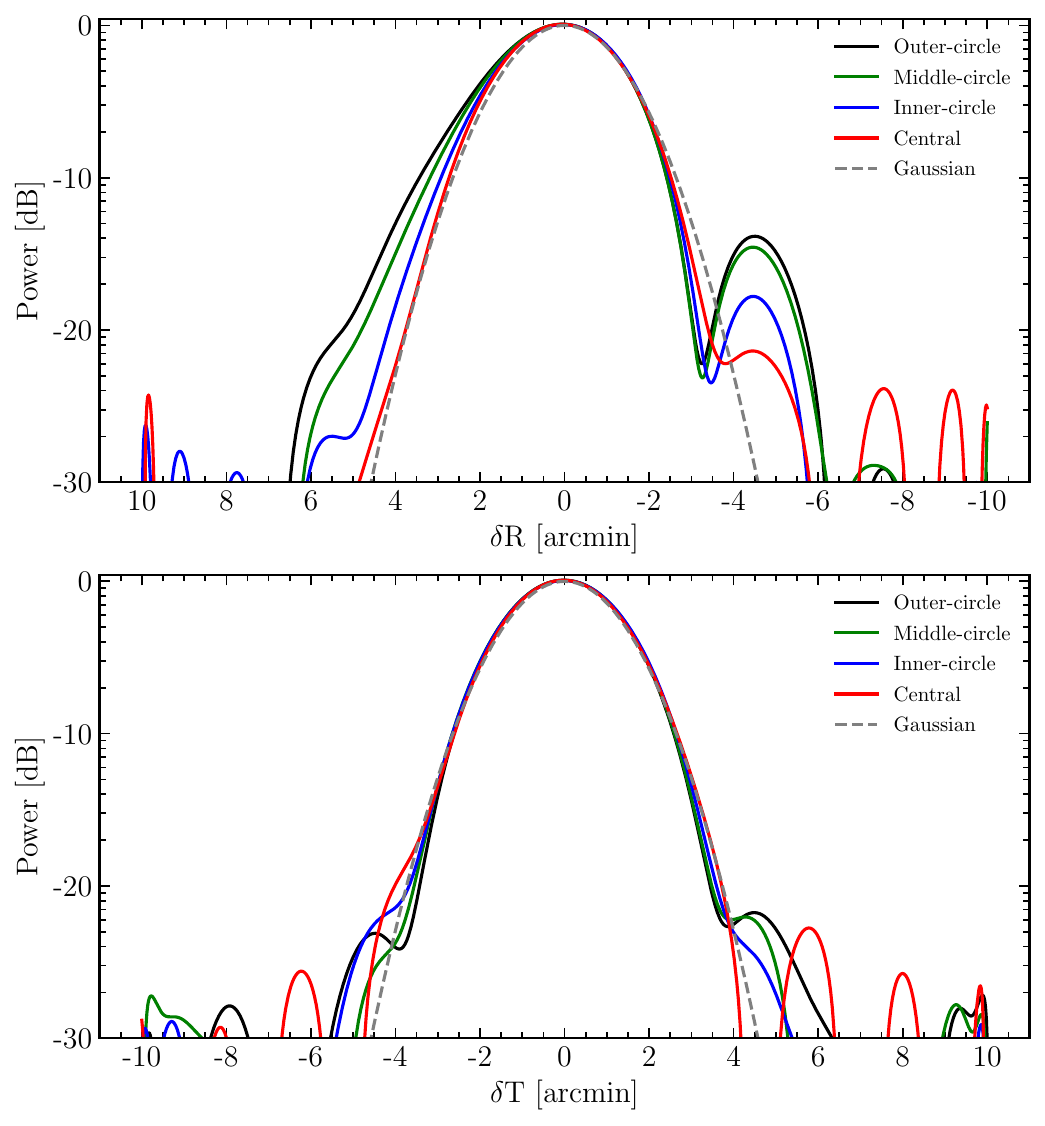}
    \caption{
    The stacked beam profiles along the radial (top panel) and tangential (bottom panel) 
    directions of the 19FA frame.
    $\delta R$ and $\delta T$ axes indicate the separation angle with respect to the beam 
    center along the radial and tangential directions, respectively.
    The beam profiles for different beam subsets are shown in different colors.
    The central beam profile is constructed by circular averaging of the center beam pattern
    and it is the same for the radial and tangential profile.
    The Gaussian profile is assumed to have $\theta_{\rm FWHM}$ is $2.9'$,
    the prior angular resolution of FAST at 1400 MHz.
    }
    \label{fig:linecut}
\end{figure}

In this work, we group the 19 feeds into four beam subsets, i.e., as shown in 
\reffg{fig: BeamPos}, the center beam, inner-circle beams, middle-circle beams, and 
outer-circle beams. To emphasize the major features of each beam subset, we 
stack the beam patterns of the same subset. 
Because the beam pattern still retains axial symmetry, with its axis of symmetry 
aligned with the radial direction of the feed array,
we rotate and stack the off-center beams in the $R-T$ coordinates, where 
$R$ and $T$ represent the radial and tangential directions of the 19FA frame.
The subset-stacked beam patterns are shown in \reffg{fig:groupstack}, where
$\delta R$ and $\delta T$ represent the separation angle with respect to the beam center
along the $R$ and $T$ axes.
We further extract the beam profile along the horizontal and vertical dashed lines
as the radial and tangential direction beam profiles.  
The radial and tangential direction beam profiles for each beam subset are shown
in the top and bottom panels of \reffg{fig:linecut}, respectively.


In \reffg{fig:linecut}, as a reference, we plot a Gaussian profile with $\theta_{\rm FWHM} = 2.9'$
in dashed gray line. 
The radial and tangential central beam profiles are the same and are constructed 
by circular averaging of the center beam pattern.
There is a significant deviation from the symmetric Gaussian beam profile for the 
off-center beams in the radial direction, while the tangential direction shows
minor deviations.


For the radial beam profile, the center beam shows several side lobes 
and its first side lobe is weak and connected with the main beam profile. 
The off-center beams display a prominent first side lobe on the side close to the 
center of the feed array.
On average, the inner-circle, middle-circle, and outer-circle beams exhibit the 
first side lobes peaked at $\delta R = -4.455' $,  $\delta R = -4.455' $, and $\delta R = -4.495' $, 
with amplitudes of $-17.79 {\rm dB}$, $-14.56 {\rm dB}$, and $-13.84 {\rm dB}$ 
with respect to the beam center, respectively. 
In addition, the middle-circle and outer-circle beams also show the second side lobes
at $\delta R = -7.317' $ and $\delta R = -7.538' $, with amplitudes of $-28.88\, {\rm dB}$ and $-29.11\, {\rm dB}$,
respectively.

\subsection{Main beam efficiency}\label{sc:beameff}

\begin{figure}

	\includegraphics[width=\columnwidth]{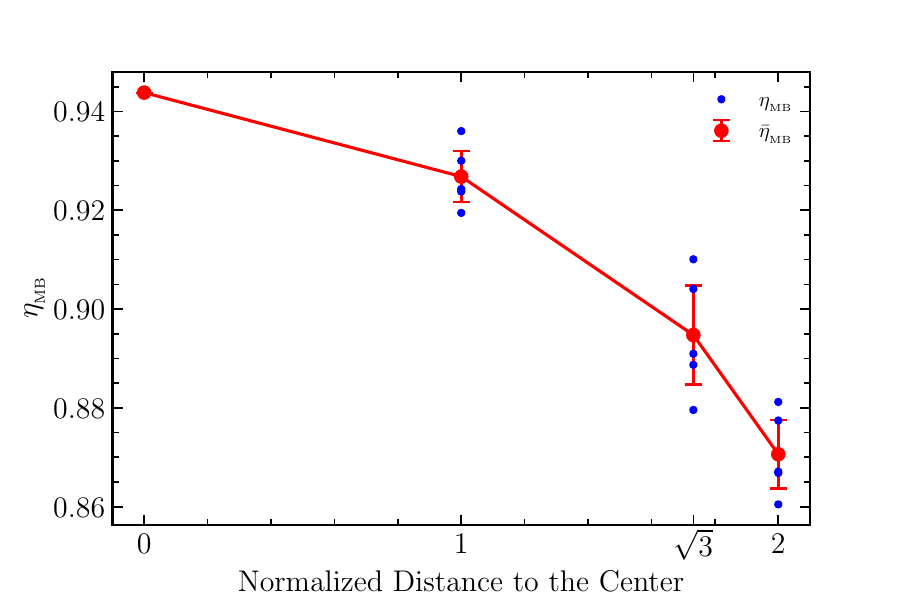}
    \caption{The main beam efficiency of 19 beams. 
    The $x$-axis represents the distance to the feed array center normalized with the 
    radius of the inner circle. The main beam efficiency of the center beam, 
    inner-circle beams, middle-circle beams, and the outer-circle beams are shown with 
    $x=0$, $1$, $\sqrt{3}$, and $2$, respectively.
    The red error bars indicate the mean and standard deviation of the main beam efficiency 
    across the beams in the same subset.
    }
    \label{fig:efficiency}
\end{figure}

The beam pattern distortion and the prominent side lobes affect the main beam efficiency.
We estimate the main beam efficiency $\eta_{\rm MB} = \Omega_{\rm MB}/\Omega_{\rm A}$, 
where $\Omega_{\rm MB}$ is the main beam solid angle and $\Omega_{\rm A}$ is the 
full beam solid angle.
According to the noise-filtered beam patterns in \reffg{fig: contour}, 
the beam patterns decline to below $0.5\%$ of the maximum beam pattern 
with $\rho \gtrsim 6'$. 
We estimate the full beam solid angle by integrating the beam profile with $\rho < 10'$.
In addition, we define the main beam size according to the center beam, i.e.,
the circular region with $\rho \leqslant \theta_{\rm FWHM}$,  
is defined as the main beam, which is close to the size of the first null. 



The main beam efficiencies of the 19 beams are shown with the blue markers in \reffg{fig:efficiency}.
The $x$-axis represents the distance to the feed array center normalized with the 
radius of the inner circle, and the beams in each subset, i.e., 
the center beam, inner-circle beams, middle-circle beams, 
and the outer-circle beams, are shown with $x=0$, $1$, $\sqrt{3}$, and $2$.
The red error bars indicate the mean and the standard deviation of the main beam efficiencies 
of the same beam subset.
The central beam has the main beam efficiency $\eta = 0.944$.
The mean main beam efficiency decreases to $0.927$, $0.895$, and $0.871$ for the  
beams in inner-circle, middle-circle, and outer-circle beam subsets, respectively. 
The decreases in the main beam efficiency for the off-center beams are 
primarily due to the growing side lobes.
However, it is corresponding to $1.7\%$, $4.9\%$, and $7.3\%$ decreases 
with respect to the center beam's efficiency, which is a minor declination.

The main beam efficiency of FAST L-band 19 feeds measured in the literature
varies, e.g. at the range $65\%$ to $85\%$ from \citet{2022SCPMA..6529705G} or 
at the range from $80\%$ to $95\%$ \citet{2021RAA....21..282S}. 
The variation is likely caused by the measurement uncertainties of the noise diode.
The value increases overall as the observation frequency increases 
and the main beam efficiency decreases as the beam moves away from the center, 
with the central beam having the maximum efficiency \citep{2022SCPMA..6529705G}.
In this research, we consider the unique beam patterns of the FAST L-band 19
feeds and use numerical integration to calculate the main beam efficiency
$\eta_{\rm MB}$. Our results are comparable with the measurement in the 
literature, in particular, consistent with \citet{2021RAA....21..282S}.

\section{Conclusion} \label{sec:conclusion}

This is the second in a series of papers focusing on intensity mapping 
in the FAST 
using drift scan surveys.
In this paper, we utilize the drift scan observation data from the 
\fathomer 
\citep{2023ApJ...954..139L}, 
as well as the data from CRAFTS 
\citep{2018IMMag..19..112L},
preprocessed with the pipeline developed in our previous work 
\citep{2023ApJ...954..139L}, 
to reconstruct the beam model for the FAST 19FA. 

We develop a beam reconstruction method based on stacking analysis. 
This approach utilizes drift-scan observational data, 
using radio continuum point sources within the drift-scan field, 
such as NVSS sources, to reconstruct the antenna’s beam pattern. 
This method minimizes the impact of beam variations caused by changes 
in antenna configuration on actual observations, thus optimizing the 
accuracy of beam measurement.

We adopt the ZP modes as the analytic basis and decompose the 
two-dimensional beam pattern using these modes. 
ZP modes with fitting coefficients below $1\%$ of the maximum coefficient are discarded to prevent overfitting. 
Specifically, $31$ ZP modes are used to construct the central beam pattern; 
on average, approximately $33$ ZP modes are used for the inner-circle beams, 
and around $44$ ZP modes for the middle- and outer-circle beams.

The center beam exhibits a circularly symmetric beam profile with suppressed 
side lobes, which can be well modeled with a Gaussian beam profile. 
For the off-center beams, the side lobes are getting more pronounced 
with the larger separation distance between the beam and the feed array center.
In addition, the side lobes become more pronounced on the side of the beam 
closer to the center of the feed array along the radial direction.
Although the circular symmetry of the off-center beam is disrupted, 
the beam pattern still retains axial symmetry, with its axis of symmetry 
aligned with the radial direction of the feed array.

We found the first side lobes averaged across the same beam subsets,
i.e. the inner-circle, middle-circle, and outer-circle beam subsets, 
peaked at $\delta R = -4.455' $,  $\delta R = -4.455' $, and $\delta R = -4.495' $, 
with amplitudes of $-17.79 {\rm dB}$, $-14.56 {\rm dB}$, and $-13.84 {\rm dB}$
with respect to the beam center, respectively. 
The corresponding main beam efficiencies slightly decrease by 
$1.7\%$, $4.9\%$, and $7.3\%$ with respect to the center beam's efficiency.

Given the current sensitivity limits, we focus exclusively on the 
Stokes I beam pattern, averaged over the frequency range of $1375$–-$1425$ MHz. 
With additional drift-scan observational data, our stacking-based beam 
reconstruction method could also facilitate studies of polarized beam patterns 
and their frequency dependence.


\section*{Acknowledgements}
This work made use of the data from FAST (Five-hundred-meter Aperture Spherical radio Telescope). 
FAST is a Chinese national mega-science facility, operated by 
National Astronomical Observatories, Chinese Academy of Sciences.
We acknowledge the support of the National SKA Program of China 
(Nos.~2022SKA0110200, 2022SKA0110202, 2022SKA0110203, 2022SKA0110100, 2022SKA0110101),
the NSFC International (Regional) Cooperation and Exchange Project (No. 12361141814),
the 111 Project (No. B16009),
and the National Natural Science Foundation of China (Nos. 12473091, 11975072, 11835009, 12473001). 
X. Sun acknowledges the support of the NSFC (grand No. 12433006).

\section*{Data Availability}
The FAST 19FA beam models presented in this paper can be accessed 
from repository \texttt{Fast19FABM}: 
\url{https://github.com/Nyarlth/Fast19FABM}.
The data underlying this article will be shared on reasonable request to the corresponding author.

\bibliography{main}{}
\bibliographystyle{aasjournal}



\end{document}